\documentclass[aps,prb,onecolumn,floats,superscriptaddress]{revtex4}
\usepackage[pdftex]{graphicx} 
\usepackage{epstopdf}
\usepackage{verbatim}
\usepackage{color}
\usepackage{subfigure}
\usepackage{tabularx}
\usepackage{amsfonts}
\usepackage{wasysym}
\usepackage{bm}
\usepackage{txfonts}
\usepackage{amssymb}
\usepackage{graphicx}
\usepackage{epstopdf}
\usepackage{multirow}
\usepackage{float}
\usepackage[linkcolor=black,anchorcolor=black,citecolor=black,hypertexnames=false]{hyperref}
\usepackage{multibib}
\renewcommand\figurename{\textbf{Figure}}
\renewcommand{\arraystretch}{1.2}
\newcommand{\bra}[1]{\ensuremath{\langle#1|}}
\newcommand{\ket}[1]{\ensuremath{|#1\rangle}}

\newcommand{\be}{\begin{equation}}
\newcommand{\bea}{\begin{eqnarray}}
\newcommand{\ee}{\end{equation}}
\newcommand{\eea}{\end{eqnarray}}

\makeatletter
\def\bib@device#1#2{}
\makeatother
\renewcommand{\arraystretch}{0}
\begin{document}

\title{Evidence for a spinon Fermi surface in a triangular lattice quantum spin liquid candidate}

\author{Yao Shen}
\affiliation{State Key Laboratory of Surface Physics and Department of Physics, Fudan University, Shanghai 200433, China}
\author{Yao-Dong Li}
\affiliation{School of Computer Science, Fudan University, Shanghai, 200433, China}
\author{Hongliang Wo}
\affiliation{State Key Laboratory of Surface Physics and Department of Physics, Fudan University, Shanghai 200433, China}
\author{Yuesheng Li}
\affiliation{Department of Physics, Renmin University of China, Beijing 100872, China}
\author{Shoudong Shen}
\affiliation{State Key Laboratory of Surface Physics and Department of Physics, Fudan University, Shanghai 200433, China}
\author{Bingying Pan}
\affiliation{State Key Laboratory of Surface Physics and Department of Physics, Fudan University, Shanghai 200433, China}
\author{Qisi Wang}
\affiliation{State Key Laboratory of Surface Physics and Department of Physics, Fudan University, Shanghai 200433, China}
\author{H. C. Walker}
\affiliation{ISIS Facility, Rutherford Appleton Laboratory, STFC, Chilton, Didcot, Oxon OX11 0QX, United Kingdom}
\author{P. Steffens}
\affiliation{Institut Laue-Langevin, 71 Avenue des Martyrs, 38042 Grenoble Cedex 9, France}
\author{M. Boehm}
\affiliation{Institut Laue-Langevin, 71 Avenue des Martyrs, 38042 Grenoble Cedex 9, France}
\author{Yiqing Hao}
\affiliation{State Key Laboratory of Surface Physics and Department of Physics, Fudan University, Shanghai 200433, China}
\author{D. L. Quintero-Castro}
\affiliation{Helmholtz-Zentrum Berlin f\"{u}r Materialien und Energie, D-14109 Berlin, Germany}
\author{L. W. Harriger}
\affiliation{NIST Center for Neutron Research, National Institute of Standards and Technology, Gaithersburg, Maryland 20899, USA}

\author{M. D. Frontzek}
\affiliation{Quantum Condensed Matter Division, Oak Ridge National Laboratory, Oak Ridge, Tennessee 37831-6393, USA}

\author{Lijie Hao}
\affiliation{Neutron Scattering Laboratory, China Institute of Atomic Energy, Beijing 102413, China}
\author{Siqin Meng}
\affiliation{Neutron Scattering Laboratory, China Institute of Atomic Energy, Beijing 102413, China}
\author{Qingming Zhang}
\affiliation{Department of Physics, Renmin University of China, Beijing 100872, China}
\author{Gang Chen$^\ast$}
\affiliation{State Key Laboratory of Surface Physics and Department of Physics, Fudan University, Shanghai 200433, China}
\affiliation{Center for Field Theory and Particle Physics, Fudan University, Shanghai, 200433, China}
\affiliation{Collaborative Innovation Center of Advanced Microstructures, Nanjing, 210093, China}
\author{Jun Zhao$^\ast$}
\affiliation{State Key Laboratory of Surface Physics and Department of Physics, Fudan University, Shanghai 200433, China}
\affiliation{Collaborative Innovation Center of Advanced Microstructures, Nanjing, 210093, China}

\begin{abstract}
\end{abstract}

\maketitle
\textbf{A quantum spin liquid is an exotic quantum state of matter
in which spins are highly entangled and remain disordered
down to zero temperature. Such a state of matter is potentially relevant to high-temperature superconductivity and quantum-information applications, and experimental identification of a quantum spin liquid state is of fundamental importance for our understanding of quantum matter. Theoretical studies have
proposed various quantum-spin-liquid ground states~\cite{Balents,XGWen,Kivelson,Anderson1973},
most of which are characterized by exotic spin excitations with
fractional quantum numbers (termed `spinon'). Here, we report
neutron scattering measurements that reveal broad spin excitations
covering a wide region of the Brillouin zone in a triangular antiferromagnet YbMgGaO$_4$. The observed diffusive spin
excitation persists at the lowest measured energy and shows a clear
upper excitation edge, which is consistent with
the particle-hole excitation of a spinon Fermi surface.
Our results therefore point to a QSL state with a spinon
Fermi surface in YbMgGaO$_4$ that has a perfect spin-1/2 triangular
lattice as in the original proposal\cite{Anderson1973} of quantum spin liquids.}

In 1973, Anderson proposed the pioneering idea
of the quantum spin liquid (QSL) in the study of
the triangular lattice Heisenberg antiferromagnet~\cite{Anderson1973}.
This idea was revived after the discovery in 1986
of high-temperature superconductivity~\cite{Anderson1987}.
A QSL, as currently understood, does not fit
into Landau's conventional paradigm of symmetry
breaking phases~\cite{PALee,XGWen,Balents,Moessner},
and is instead an exotic state of matter
characterized by spinon excitations and emergent
gauge structures~\cite{Kivelson,XGWen,Balents,Moessner}.
The search for QSLs in models and materials~\cite{organics1,organics2,kappaET,cscucl,YoungLee}
has been partly facilitated by the Oshikawa-Hastings-Lieb-Schultz-Mattis (OHLSM)
theorem that may hint at the possibility of QSLs in Mott insulators with odd
electron fillings and a global U(1) spin rotational
symmetry~\cite{Oshikawa,Hastings,LSM}.
Indeed, a continuum of spin excitations has been observed
in a kagome-lattice material ZnCu$_3$(OD)$_6$Cl$_2$ (refs ~\onlinecite{YoungLee,punk}).
However, the requirement of the U(1) spin rotational symmetry,
prevents the application of OHLSM theorem in
strong spin-orbit-coupled (SOC) Mott insulators in which the
spin rotational symmetry is completely absent.
A recent theory addressed this limitation of the OHLSM theorem,
arguing that, as long as time-reversal symmetry is preserved,
the ground state of an SOC Mott insulator
with odd electron fillings must be exotic~\cite{Ashvin}.

The newly discovered triangular antiferromagnet
YbMgGaO$_4$ (refs~\onlinecite{Yuesheng1,Yuesheng2})
displays no indication of magnetic ordering or symmetry breaking
at temperatures as low as 30 mK despite approximately 4 K
for the spin interaction energy scale.
Because of the strong SOC of the Yb electrons, YbMgGaO$_4$ was the first QSL
to be proposed beyond the OHLSM theorem~\cite{Yuesheng2}.
The thirteen $4f$ electrons of the Yb$^{3+}$ ion form
the spin-orbit-entangled Kramers doublets
that are split by the D$_{3d}$ crystal electric fields~\cite{Yaodong,ross,Yaodong2}.
At temperatures considerably lower than the crystal field gap ($\sim 420$ K),
the magnetic properties are captured by the ground state
doublet described by an effective spin-1/2 local moment.
This is further confirmed by a measured magnetic entropy
equal to R$\ln 2$ per Yb$^{3+}$ ion~\cite{Yuesheng1}.
Figure 1a,b shows that the YbO$_6$ octahedra form
well separated triangular layers. Because of the large chemical
difference between Yb$^{3+}$
and the nonmagnetic Mg$^{2+}$/Ga$^{3+}$ ions, intra-triangular-layer impurities
are prevented in YbMgGaO$_4$ (refs~\onlinecite{Yuesheng1,Yuesheng2,Yaodong2}).
Hence, the Yb system is the perfect spin-1/2 triangular lattice
antiferromagnet appearing in the original proposal outlined by Anderson~\cite{Anderson1973}.

To characterize the Yb local moment behavior,
we first measured the magnetic susceptibility of a single-crystalline
YbMgGaO$_4$ (Fig. 1c). For both magnetic fields applied parallel to and normal to the $c$ axis,
we found predominant antiferromagnetic spin interactions that were evidenced by
 negative Curie-Weiss temperatures (Fig. 1d).
Because of the anisotropy of the spin interaction,
the Curie-Weiss temperatures for $H \perp c$ and $H \parallel c$ were not identical,
with $\Theta_{\text{CW},\perp} = -4.78$ K and $\Theta_{\text{CW},\parallel} = -3.2$ K, consistent with previous measurements~\cite{Yuesheng1,Yuesheng2}.
We examined the magnetic susceptibilities in field cooling (FC)
and zero-field cooling (ZFC) measurements.
No splitting was detected between the FC and ZFC results down to 2 K,
indicating the absence of spin glassy transitions (Fig. 1c).

The Curie-Weiss temperature and the spin excitation bandwidth
(discussed subsequently) set the energy scale for the spin interactions.
Our elastic neutron scattering measurements revealed no
magnetic Bragg peaks (Extended Data Fig. 2) at temperatures as low as 30 mK,
considerably lower than the Curie-Weiss temperature($\sim4$ K)
and spin excitation bandwidth($\sim$ 17 K); this is consistent
with previous specific heat and susceptibility measurements.
To reveal the intrinsic quantum dynamics of the Yb local moments,
we used inelastic neutron scattering (INS) to study the spin
excitations in YbMgGaO$_4$ single crystals at $\sim$70 mK.
Constant energy images are presented in Fig. 2a-e, indicating diffusive magnetic excitations for all measured energies.
The scattering spectral weights are spread broadly
in the Brillouin zone, whereas the spectral intensities near the
zone center ({\sl i.e.}, the $\Gamma$ point) were suppressed.
For a low energy transfer of 0.3 meV, the spectral intensity
was slightly more pronounced around the M points, while the broad
continuum across the Brillouin zone still carried the vast majority of the spectral weight (Fig. 2a).

Figure 3a displays the contour plot of spectral intensity
along the high symmetry momentum directions,
namely M-K-$\Gamma$-M-$\Gamma$, in $E$-$\bf{Q}$ space.
Similar to the constant energy images shown in Fig. 2a-e,
the spectral intensity was broadly distributed in
momentum for all energies measured. Moreover, a clear
V-shaped upper bound of the excitation energy
was found near the $\Gamma$ point (Fig. 3a, dotted line).
The spin excitation intensity gradually decreased with
increasing energy and vanished above approximately 1.5 meV.
This feature is further confirmed by the $\bf{Q}$ scans
in Fig. 4a, b as well as the $E$ scans
at a few given momentum points ($\Gamma$, M, K) in Fig. 4c.

The broad continuum is an immediate consequence and
strong evidence of spinon excitations in QSLs~\cite{YoungLee,PALee,Balents}. This differs from magnon-like excitations that would peak
strongly at specific momenta in the reciprocal space,
with or without static magnetic order\cite{KFS,Wang}.
In general, the spinful excitations in QSLs are
carried by deconfined spinons~\cite{PALee,Balents}.
For most experimentally relevant QSLs, the spinons carry half-integer spins.
One neutron spin flip event in an INS
measurement creates an integer spin change that necessarily
excites two (or more) spinons~\cite{Balents}.
Therefore, the energy transfer, $E$, and the momentum transfer, ${\bf p}$,
of the neutron are shared by two spinon excitations that are
created by the neutron spin flip. According to energy-momentum conservation,
we have $E({\bf p}) = \omega_s ({\bf k}) + \omega_s ({{\bf p} - {\bf k}})$,
where $\omega_s ({\bf k})$ is the spinon dispersion.
This relation implies the presence of an excitation
continuum in the INS spectrum. The broad continuums in
Fig. 2a-e and Fig. 3a at different energies are as expected
for the continuum excitations of two spinons.

The broad neutron-scattering spectral intensity that persists to the lowest
energy measured suggests a high density of spinon scattering states at low energies.
This cannot be explained by a Dirac QSL~\cite{KagomePRL,XGWen},
in which the spectral continuum at low energies would
concentrate near a few discrete momenta
that connect the Dirac cones (Methods),
or any simple gapped QSL. Because of the gap,
the spectral intensity would exceed a specified energy threshold.
Even if the gap is smaller than the lowest measured energy,
except for special reasons~\cite{punk} the spinon excitations would simply occupy one or
a few discrete spots in reciprocal space at low energies~\cite{Sachdev},
gradually expanding with increasing energy, rather
than the observed broad continuum at all energies
and the diminishing spectral weight at $\Gamma$ (Fig. 3a).
Moreover, both the Dirac QSL and gapped QSLs are inconsistent
with the low-temperature sublinear power-law behavior
of the heat capacity~\cite{Yuesheng2}.
In contrast, the spinon Fermi surface QSL, with a high density of spinon
states near the spinon Fermi surface, provides a consistent explanation
for the INS results of YbMgGaO$_4$.

To account for these possible QSL signatures in YbMgGaO$_4$, we consider a minimal
mean-field spinon Hamiltonian with a uniform spinon hopping
on the triangular lattice. With a zero background flux for the spinons,
the spinons form a large Fermi surface in the Brillouin zone (Fig. 2g).
Although the anisotropic spin exchange caused by the SOC~\cite{Yuesheng2,Yaodong,Yaodong2}
breaks the spin symmetry of this simple model, the mean-field state
considered here captures the essential properties of the spinon
Fermi surface QSL in this system. For this spinon
Fermi surface state, one neutron spin flip excites
one spinon particle-hole pair across the Fermi surface.
Therefore, the dynamic spin structure factor
$\mathcal{S} ({{\bf p}, E})$,
measured by the INS, directly probes the spinon
particle-hole excitations across the spinon Fermi surface (Fig. 2g).

For a low $E$, a minimum momentum transfer $p_{\text{min}}\approx E/v_{\text F}$
is required to excite the spinon particle-hole pairs,
where $v_{\text F}$ is the Fermi velocity. Therefore, the
spectral intensity near the $\Gamma$ point
should be gradually suppressed with increasing energy,
leading to an upper bound of the excitation energy near
the $\Gamma$ point (Extended Data Fig. 5b), which is clearly observed in Fig. 3a (V-shaped dotted line).
For a typical finite $E$, the calculated spectrum based on
the spinon particle-hole continuum is shown in Fig. 2f.
This is qualitatively consistent with the
experimental observation of the broad spinon continuum
in the reciprocal space. Finally, when $E$ exceeds the spinon bandwidth,
the single spinon particle-hole excitation process is suppressed,
and the spinon excitation intensity is suppressed accordingly (Extended Data Fig. 5b).
This feature is consistent with the vanishing of the spectral
intensity above $\sim 1.5$ meV (dotted line) in Fig. 3a and Fig. 4c.
Therefore, we propose that YbMgGaO$_4$ is a QSL with a spinon Fermi surface.

The spinon Fermi surface alone has a constant density of states and
would give a linear-$T$ heat capacity. To account for the
$C_v \approx \text{Constant}\times T^{2/3}$ behavior in YbMgGaO$_4$~\cite{Yuesheng2},
we further propose the candidate QSL as the spinon Fermi surface U(1) QSL
where the strong U(1) gauge fluctuation invokes a self-energy correction
in the spinons, thus enhancing the low-energy density of
states~\cite{PALeeSungSik,Motrunich,LeeNagaosa}.

Finally, during the review  of our paper, a related preprint\cite{paddison} appeared discussing the role of the next-nearest-neighbor interactions in the formation of the QSL state in YbMgGaO$_4$.

\textbf{Acknowledgements}

We thank Dunghai Lee, Shiyan Li, Yuanming Lu, Xiaoqun Wang, and
especially Jia-Wei Mei for useful discussions, and Fengqi Song for the assistance in the magnetic susceptibility measurements.
This work was supported by the National Key R\&D Program of the MOST of China (Grant No. 2016YFA0300203), the Ministry of Science and
Technology of China (Program 973: 2015CB921302), and
the National Natural Science Foundation of China (Grant No. 91421106).
Yao-Dong Li and Gang Chen were supported by the Thousand Youth Talent
Program of the People's Republic of China. This research used resources at the High Flux Isotope Reactor, a DOE Office of Science User Facility operated by the Oak Ridge National Laboratory.

\textbf{Author information}

Correspondence and requests for materials should be
addressed to J.Z. (zhaoj@fudan.edu.cn) or G.C. (gchen\_physics@fudan.edu.cn).

\newpage

\begin{figure}[!ht]
\includegraphics{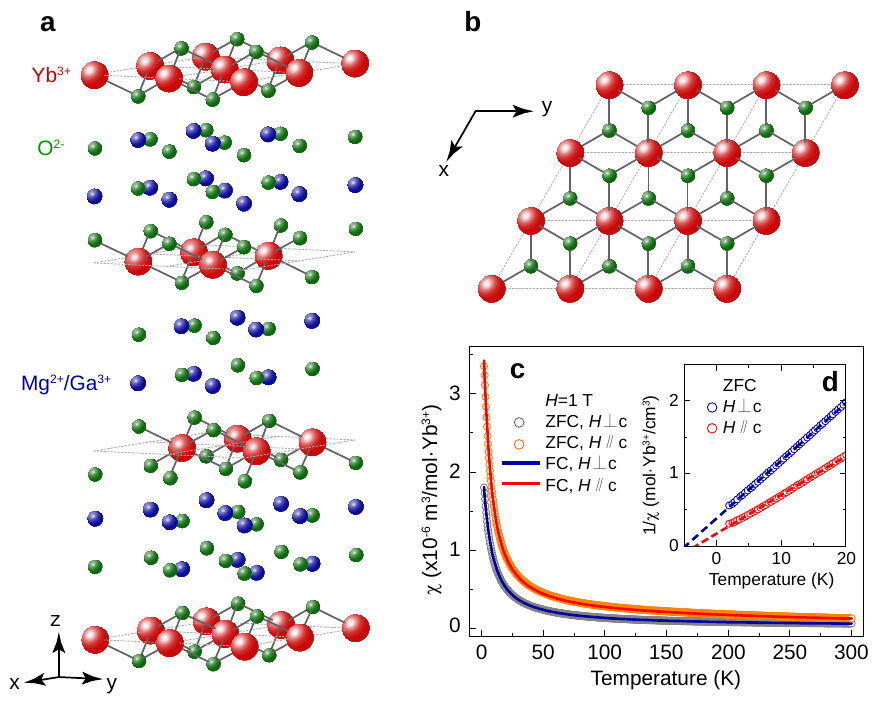}
\caption{ \textbf{Crystal structure and magnetic susceptibility of YbMgGaO$_4$ single crystals}. \textbf{a,b,} Schematic of the YbMgGaO$_4$ crystal structure. The dashed line indicates the unit cell. \textbf{c,} DC magnetic susceptibility measured under zero-field cooling (ZFC) and field cooling (FC) on YbMgGaO$_4$ single crystals. Paramagnetic behavior is observed at low temperature with no obvious differences between ZFC and FC data. \textbf{d,} Inverse susceptibility at low temperature ($\le$ 20 K) fitted with the Curie-Weiss law (dashed line). The fitting results in Curie temperatures of $\Theta_W = -4.78$ K and -3.2 K for magnetic fields (\textit{H} $= 1$ T) perpendicular and parallel to the \textit{c} axis, respectively.
}
\end{figure}

\newpage

\begin{figure}[!ht]
\includegraphics{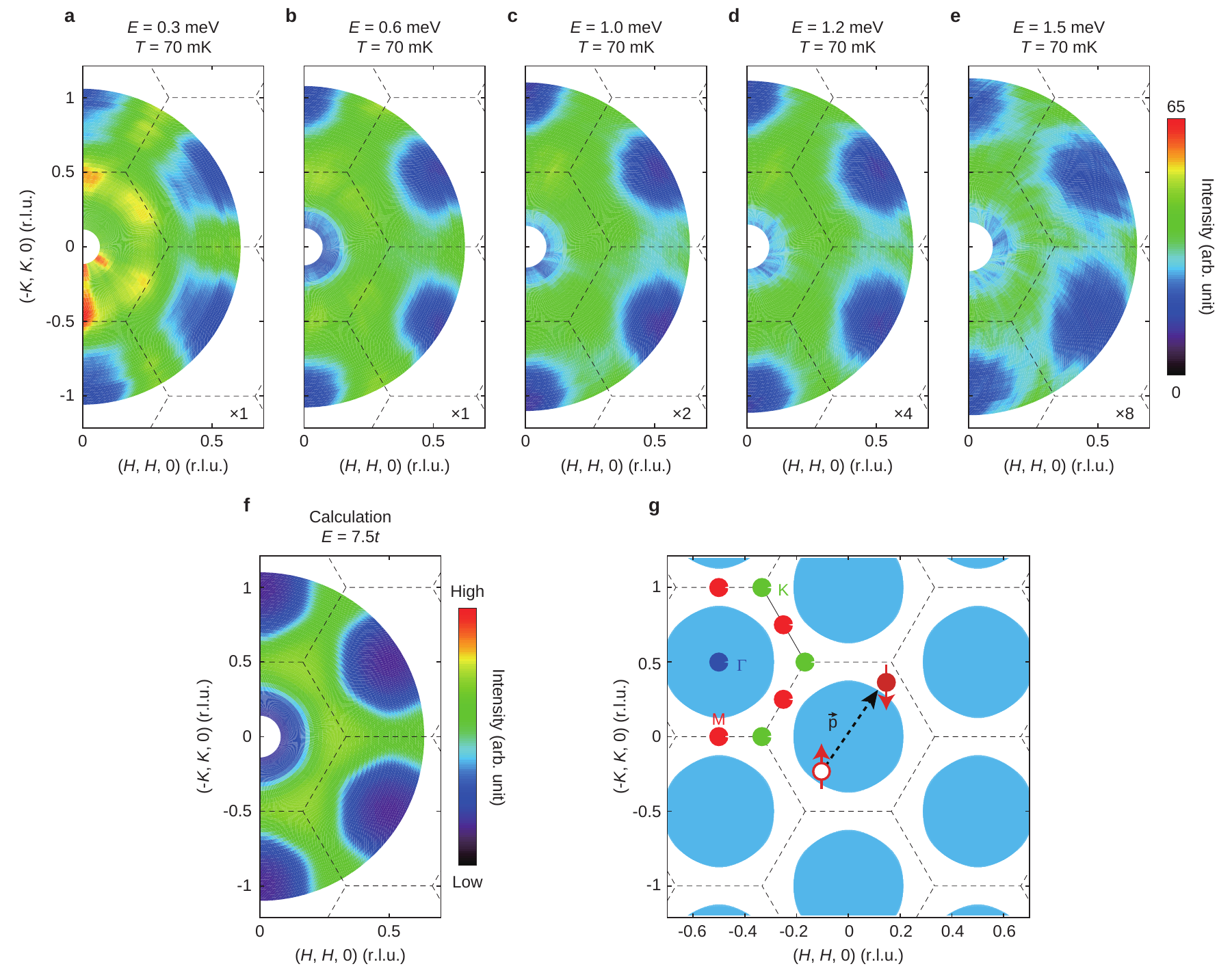}
\caption{ \textbf{Measured and calculated momentum dependence of the spin excitations, and calculated spinon Fermi surface of YbMgGaO$_4$ at 70 mK}. \textbf{a-e,} Constant energy image at indicated energies, displaying diffusive magnetic excitations covering a wide region of the Brillouin zone. The scattering intensity at different energies has been multiplied by a scale factor for clarity, noted in the bottom right corners of the panels. The intensity is represented by a linear color scale throughout the paper. The data were collected on ThALES using the Flatcone detector, and were corrected for neutron beam self-attenuation (Methods). \textbf{f,} Calculated momentum dependence of the spin excitations for a typical finite $E$. Here $t$ is the hopping amplitude between nearest-neighbor sites. \textbf{g,} Spinon Fermi surface calculated using the model described in the main text. The black arrow indicates a spinon particle-hole excitation, and dashed lines indicate the Brillouin zone boundaries of the conventional unit cell ($\textit{a} = \textit{b} = 3.40 {\AA}$, $\textit{c} = 25.12 {\AA}$). High symmetry points M, K and $\Gamma$ are labelled by red, green, and blue dots, respectively.
}
\end{figure}

\newpage

\begin{figure}[!ht]
\includegraphics{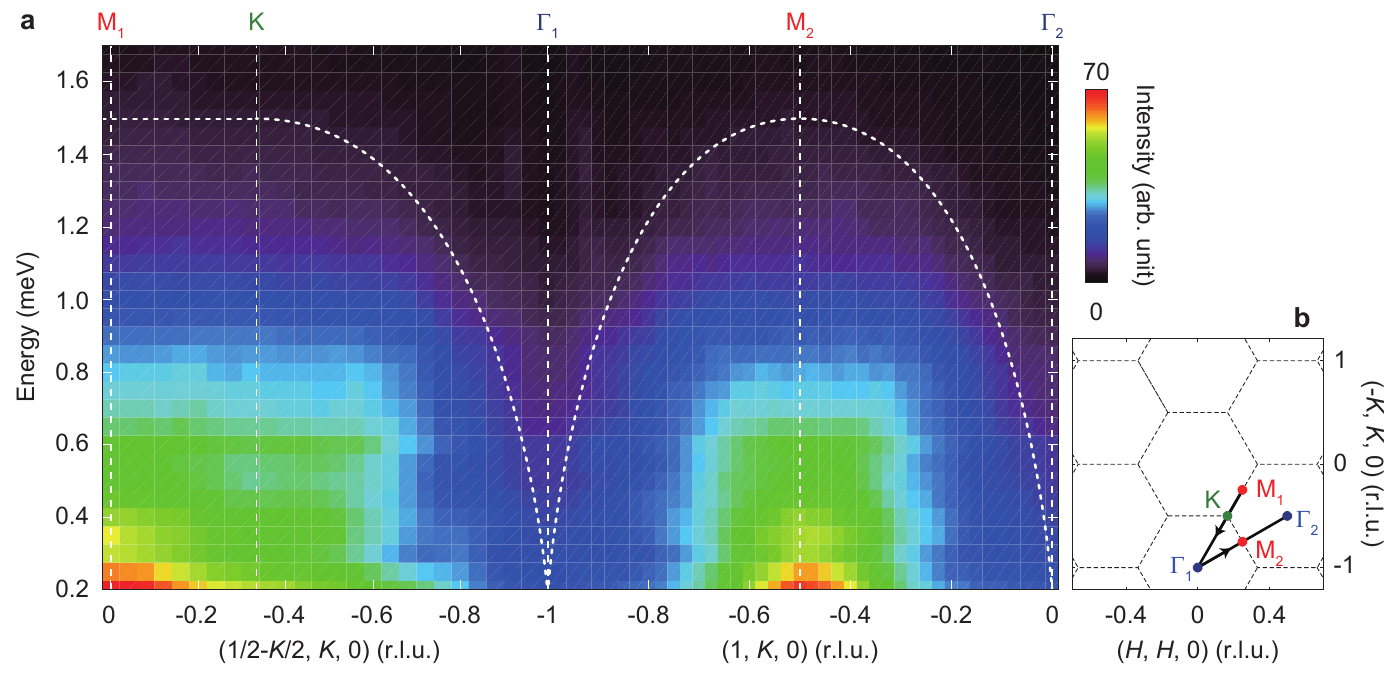}
\caption{ \textbf{Intensity contour plot of spin excitation spectrum along the high-symmetry momentum directions}. \textbf{a,} Intensity contour plot along the (1/2-\textit{K}/2, \textit{K}, 0) and (1, \textit{K}, 0) directions as illustrated in \textbf{b}. Vertical dashed lines represent the high-symmetry points, and dotted lines indicate the upper bounds of spin excitation energy. \textbf{b,} Sketch of reciprocal space. Dashed lines indicate the Brillouin zone boundaries.
}
\end{figure}

\newpage

\begin{figure}[!ht]
\includegraphics{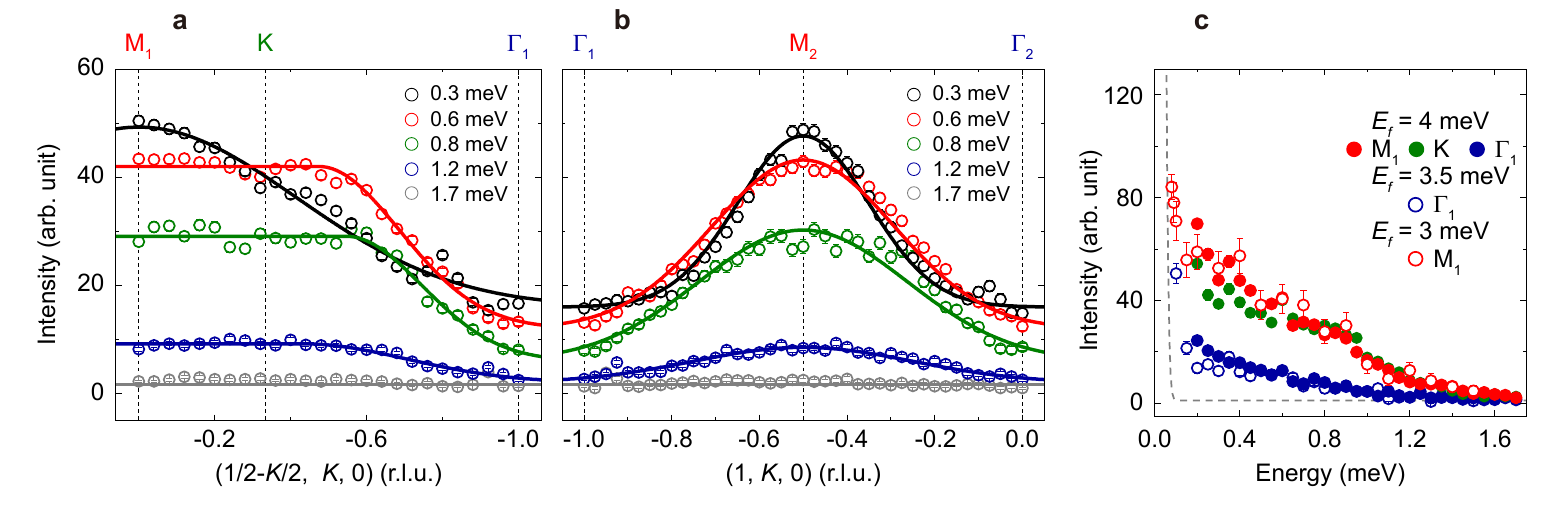}
\caption{ \textbf{Constant energy scans along the symmetry directions and constant $\bf{Q}$ scans at the high-symmetry points}. \textbf{a,b,} Constant energy scans along the (1/2-\textit{K}/2, \textit{K}, 0) and (1, \textit{K}, 0) directions. The solid lines are guides to the eye. \textbf{c,} Constant $\bf{Q}$ scans at M, K, and $\Gamma$ points with the final energy fixed at $E_f$ = 3, 3.5 and 4 meV. The sharp upturn of the scattering below $\sim 0.1$ meV is due to contamination from incoherent elastic scattering at $E=0$ meV (dashed line, $E_f$=3 meV). Error bars, 1 s.d.
}
\end{figure}

\newpage

\textbf{Methods}
\\
\textbf{Sample growth and characterizations.}

High-quality YbMgGaO$_4$ single crystals were synthesized using the optical floating zone technique \cite{Yuesheng2}. A representative single crystal which is optically transparent with mirror-like cleaved surfaces is shown in Extended Data Fig. 1a. Our X-ray diffraction (XRD) measurements revealed that all reflections from the cleaved surface could be indexed by (0, 0, L) peaks of triangular YbMgGaO$_4$, and no impurity phases were observed (Extended Data Fig. 1b). The full width at half maximum (FWHM) of the rocking curve of the (0, 0, 18) peak was $\sim0.009^\circ$, indicating an extremely high crystallization quality (Extended Data Fig. 1c). This was further confirmed by the sharp and clear diffraction spots in the X-ray Laue pattern (Extended Data Fig. 1d). Powder XRD pattern on ground single crystals also revealed no indication of impurity phases (Extended Data Fig. 1e). The Rietveld refinements\cite{FULLPROF} confirm that the XRD pattern can be described by the $R\overline{3}m$ space group. The refined structural parameters are given in Extended Data Table 1. These results suggested that the YbMgGaO$_4$ single crystal possessed a perfect triangular lattice with no detectable impurities. This is consistent with previous measurements that have demonstrated that the impurity/isolated spins are less than 0.04\% in similar samples\cite{Yuesheng1,Yuesheng2}. Although the Mg/Ga site disorder in the non-magnetic layers does not directly impact the exchange interaction between the Yb local moments, it may have an indirect effect and could lead to some exchange disorder. It seems that this disorder is not significant as no signs of spin freezing were observed. In general, a quantum spin liquid is a robust quantum state that is stable against any local perturbation if the perturbation is not too strong. Therefore, if a quantum spin liquid is realized as the ground state for YbMgGaO$_4$, the possible exchange disorder will not destabilize this state if the disorder strength is not significant.

In addition, the field dependence of magnetization in our single crystal displayed a linear behavior above 12 T (Extended Data Fig. 1f), indicative of a fully polarized state. The Van Vleck susceptibility extracted from the linear-field-dependent magnetization data was subtracted in Fig. 1d.

\textbf{Neutron scattering experiments.}

Inelastic neutron scattering measurements were carried out on the ThALES cold triple-axis spectrometer at the Institut Laue-Langevin, Grenoble, France and the FLEXX cold triple-axis spectrometer in the BER-II reactor at Helmholtz-Zentrum Berlin, Germany\cite{FLEXX}. For the ThALES experiment, silicon (111) was used as a monochromator and analyzer; the final neutron energies were fixed at \textit{E}$_f = 3$ meV (energy resolution $\sim 0.05$ meV), \textit{E}$_f = 3.5$ meV (energy resolution $\sim 0.08$ meV) or \textit{E}$_f = 4$ meV (energy resolution $\sim 0.1$ meV). For the FLEXX experiment, pyrolythic graphite (002) was used as a monochromator and analyzer. Contamination from higher-order neutrons was eliminated through a velocity selector installed in the front of the monochromator. The final neutron energy was fixed at \textit{E}$_f = 3.5$ meV (energy resolution $\sim0.09$ meV). Three (six) pieces of single crystals with total a mass of $\sim 5$ g ($\sim 19$ g) were coaligned in the (HK0) scattering plane for the ThALES (FLEXX) experiment. The FWHM of the rocking curve of the coaligned crystals for the ThALES and FLEXX experiments were $\sim0.95^\circ$ and $\sim0.92^\circ$, respectively. The elastic neutron scattering experiment were carried out at WAND neutron diffractormeter at the High Flux Isotope Reactor, Oak  Ridge  National  Laboratory, U.S.A.. For the low temperature experiments, a dilution insert for the standard $^4$He cryostat was used to reach temperatures down to $\sim30-70$ mK.

Because of the nonuniform shape of the single crystal, the relatively large sample volume, and the extremely broad spin-excitation spectrum, the neutron beam self-attenuation (by the sample) may require consideration. In most cases the self-attenuation is dependent on only the distance traversed by the neutrons through the sample. We observed the self-attenuation effect in an elastic incoherent scattering image of our sample at 20 K, which exhibited an anisotropic intensity distribution (Extended Data Fig. 3a). The self-attenuation effect was also observed in the raw constant energy images (Extended Data Fig. 3b-f), which were shown to be anisotropic, with slightly higher intensities occurring at approximately the same direction as that observed in the elastic incoherent scattering images. The self-attenuation can be corrected by normalizing the data with the elastic incoherent scattering image, i.e., the elastic incoherent scattering intensity, which is dependent on the sample position ($\omega$) and scattering angle ($2\theta$), is converted to a linear attenuation correction factor for the scattering images measured at different energies. The normalized constant-energy images are presented in Fig. 2a-e, revealing a nearly isotropic intensity distribution.

Extended Data Figure 4 shows the spin excitation spectrum at 20 K, which is broadened and weakened compared with that at 70 mK (discussed subsequently).

\textbf{Spinon Fermi surface and dynamic spin structure factor.}

Here we explain the spinon mean-field state that is employed to explain
the dynamic spin structure factor of the neutron scattering experiments.
As we have proposed in the main text, the quantum spin liquid (QSL) with
a spinon Fermi surface gives a compatible explanation for the inelastic
neutron scattering result of YbMgGaO$_4$.

To describe the candidate spinon Fermi surface QSL state in YbMgGaO$_4$,
we formally express the Yb$^{3+}$ effective spin as the bilinear of
the fermionic spinon with ${\bf S}_i = \sum_{\alpha\beta}
\frac{1}{2} f^{\dagger}_{i\alpha} \boldsymbol{\sigma}_{\alpha\beta}
f^{\phantom\dagger}_{i\beta}$ with a Hilbert space constraint
$\sum_{\alpha} f^{\dagger}_{i\alpha} f^{\phantom\dagger}_{i\alpha} = 1$,
where $\boldsymbol{\sigma}$'s are the Pauli matrices,
and $f^\dagger_{i \alpha}$ creates a spinon with spin $\alpha$ at
site $i$, and $\alpha = \uparrow, \downarrow$.
For the QSL with a spinon Fermi surface, we propose a minimal
mean-field Hamiltonian for the spinons on the triangular lattice.
We consider a uniform spinon hopping with a zero background flux,
\begin{equation}
H_{\text{MFT}} = - t \sum_{ \langle {ij} \rangle }
( f^\dagger_{i \alpha} f^{}_{j \alpha} + h.c.)
- \mu \sum_i f^\dagger_{i \alpha} f^{}_{i \alpha},
\label{spinmft}
\end{equation}
where $t$ is the mean-field parameter, and represents
the hopping amplitude between nearest-neighbor sites.
The chemical potential $\mu$ is to impose the
Hilbert space constraint on average. Here, we have treated
the spinons freely by neglecting the gauge fluctuations.
This mean-field state simply gives a single spinon dispersion,
\begin{equation}
\omega_{\bf k} = - t \big[ \sum_{\{ {\bf a}_i \} }
\cos ({\bf k}\cdot {\bf a}_i) \big] - \mu,
\end{equation}
where $\{ {\bf a}_i \}$ are six nearest-neighbor vectors
of the triangular lattice. Due to the Hilbert space constraint,
the spinon band is half-filled, leading to a large Fermi
surface in the Brioullin zone (Extended Data Fig. 5a).

Inelastic neutron scattering measures the dynamic spin structure factor,
\bea
\mathcal{S} ({{\bf p}, E} )
	&=& \frac{1}{N} \sum_{i,j}  e^{i {\bf p}
	\cdot ({\bf r}_i - {\bf r}_j)} \int dt e^{-iEt}
	\langle {\bf S}^-_i (t) \cdot {\bf S}^+_j (0) \rangle \nonumber \\
	&=& \sum_n \delta \Big(E-(E_n ({\bf p}) - E_0) \Big)
	\left| \bra{n} \,{\bf S}^+_{\bf p}\, \ket{\Omega} \right| ^2
	\label{7}
\eea
where $N$ is total number of lattice sites,
the summation goes over all eigenstates, and $\ket{\Omega}$ refers
to the spinon ground state with the spinons filling
the Fermi sea. Here $E_0$ is the energy of the ground state,
and $E_n ({\bf p})$ is the energy of the $n$-th
excited state with momentum ${\bf p}$. In the actual calculation,
due to the energy resolution of the experiments,
the delta function is taken to have a broadening with
$\delta(E-\epsilon) = \frac{\eta/\pi}{(E-\epsilon)^2+\eta^2}$.
Since ${\bf S}^{+}_{\bf p} = \sum_{\bf k}
f^\dagger_{{\bf k+p}\uparrow} f^{}_{{\bf k}\downarrow}$,
the summation in Eq.(\ref{7}) would be over all possible
spin-1 excited states that are characterized by one spinon
particle-hole pair across the spinon Fermi surface (Fig. 2g)
with a total momentum ${\bf p}$ and a total energy $E$.
As we show in Fig. 2f and Extended Data Fig. 5b and have discussed
in the main text, this spinon Fermi surface
QSL state gives the three crucial
features of the inelastic neutron scattering results:
1) The broad continuum that covers the large portion of the Brioullin zone.
2) The broad continuum persisting from the lowest energy transfer to the
highest energy transfer.
3) The clear upper excitation edge near the $\Gamma$ point.

In our calculation of Fig. 2f and Extended Data Fig. 5b, we choose the lattice
size to be $40 \times 40$ and $\eta = 1.2t$, in accordance with
the energy and momentum resolution of the instruments.
The energy scale of Fig. 2f is set to be $7.5t$.

Here we explain the details of the dynamic spin structure factor
in  Fig. 2f and Extended Data Fig. 5b based on the particle-hole excitation of the spinon Fermi surface.
For an infinitesimal energy transfer, the neutrons simply probe the spinon Fermi surface.
As the spinon particle and hole can be excited anywhere near the Fermi surface,
the neutron spectral intensity appears from ${\bf p}=0$ to ${\bf p} = 2 {\bf k}_{\text F}$,
where ${\bf k}_{\text F}$ is the Fermi wavevector. Since $|2{\bf k}_{\text F}|$ already
exceeds the first Brillouin zone, the neutron spectral intensity then covers
the whole Brillouin zone including the $\Gamma$ point. For a small but finite $E$,
as we explain in the main text, it requires a minimal momentum
transfer $p_{\text{min}} \approx E/v_{\text F}$ to excite the spinon particle-hole pairs.
Therefore, the spectral intensity gradually moves away from the $\Gamma$ point
as $E$ increases. Because it is always possible to excite the spinon particle-hole pair
with the momenta near the zone boundary, the spectral intensity is not much effected
at the zone boundary as $E$ increases. Thus, the broad continuum continues to cover
a large portion of the Brillouin zone at a finite $E$.

With the free spinon mean-field model $H_{\text{MFT}}$, we further calculate the spectral weight along the energy direction for fixed momenta. The discrepancy between the theoretical results in Extended Data Fig. 5d and the experimental results in Extended Data Fig. 5e occurs at low energies. We attribute this low-energy discrepancy to the limitation of the free spinon theory that ignores the gauge fluctuations. The enhancement of the low energy spectral weight compared to the free spinon results is then identified as the possible evidence of the strong gauge fluctuations in the system which we elaborate in the following discussion of the heat capacity behaviors.

To account for the heat capacity behavior, we have suggested that the
candidate QSL is a spinon Fermi surface U(1) QSL in the main text.
Here we elaborate on this point and discuss the U(1) gauge fluctuation of this state in detail.
The stability of the U(1) QSL with a spinon Fermi surface
against the spinon confinement has been addressed extensively
in the literature~\cite{LeeSS1,Hermele}. It was proposed and understood
that the large densities of gapless fermionic spinons on the spinon
Fermi surface help suppress the instanton events of the compact
U(1) gauge field in two-dimensional U(1) QSL~\cite{LeeSS1,Hermele}.
The proliferation of the instanton events is the cause of the gauge
confinement of a U(1) lattice gauge theory for a U(1) QSL without
gapless spinons~\cite{Polyakov}. Since the instanton event is
suppressed here, the compactness of the U(1) gauge field is no longer an issue, and
the low-energy property of our U(1) QSL is then described by gapless
fermionic spinons coupled with a noncompact U(1) gauge field~\cite{PALeeSungSik,LeeSS1,LeeSS2}.
Due to the coupling to the gapless spinons, the U(1) gauge photon is
overly Landau damped and becomes very soft. The soft gauge photon further
scatters the fermionic spinons strongly, gives a self-energy correction
to the spinon Green's function, and kills the spinon quasi-particle weight~\cite{PALeeSungSik,LeeNagaosa,LeeSS1,LeeSS2}. The resulting spinon non-Fermi liquid state
has an enhanced density of low energy spinon states that gives a sublinear
power law temperature dependence for the low-temperature
heat capacity~\cite{PALeeSungSik,LeeNagaosa,LeeSS1,LeeSS2}. In addition to the heat capacity behavior, we find that, the enhanced density of the low energy spinon states, due to the spinon-gauge coupling and the U(1) gauge fluctuation, is consistent with the enhanced spectral intensities at low energies for the fixed momenta in Extended Data Fig. 5e.

The stability of the spinon non-Fermi liquid against spinon pairing
has also been considered theoretically~\cite{Max}.
When the spinons pair up just like the Cooper pairing of electrons in a superconductor,
the continuous part of the U(1) gauge field becomes massive due to
the Anderson-Higgs' mechanism, leaving the Z$_2$ part of the gauge
field unaffected. The resulting state
from the spinon pairing of a spinon Fermi surface U(1) quantum spin liquid
is a Z$_2$ QSL.
Such a spinon pairing scenario was proposed to account
the very low temperature behaviors of the $\kappa$-(BEDT-TTF)$_2$Cu$_2$(CN)$_3$
organic spin liquid~\cite{KappaET}.
For YbMgGaO$_4$, however, we do not find any evidence of spinon pairings in
either thermodynamic or spectroscopic measurements.
While the INS measurement might be constrained by the energy resolution,
the thermodynamic measurement did not find any suppression of the density of states
down to the lowest temperatures~\cite{Yuesheng1,Yuesheng2}.
If the spinon pairing instability may occur for YbMgGaO$_4$,
it can only occur at a much lower temperature or energy scale
than the current and previous experimental temperatures~\cite{Yuesheng1,Yuesheng2}.
In any case, the presence of a spinon Fermi surface
is the precondition for any spinon pairing instability.

We now discuss the finite-temperature thermal effect of the QSL. For the spinon Fermi surface U(1) QSL in 2D, there is no line-like object in the excitation spectrum. Therefore, as one increases the temperatures from this QSL ground state, there is no thermal phase transition by proliferating any extended line-like excitations. Moreover, from the symmetry point of view, the spinon Fermi surface U(1) QSL is not characterized by any symmetry. So there is no symmetry breaking transition as the temperature is increased. The absence of the thermal phase transition is consistent with what has been observed in YbMgGaO$_4$. As one increases the temperature from the $T = 0$ K QSL ground state, the system involves more thermal superposition of excited states and gradually loses its quantum coherence. 20 K is approximately the energy scale of the spin excitation bandwidth, which sets the interaction energy scale between the Yb local moments. At this temperature, the correlation between the local moments cannot be ignored. Its consequence is the diffusive feature in the inelastic neutron scattering spectrum. This is consistent with our data measured at 20 K in Extended Data Fig. 4a, b, where the spectral weight becomes more diffusive.

Finally, we comment on the weak spectral peak at the M points at low
energies (Fig. 2a, 3a).
This non-generic feature of the neutron spectrum,
however, is not obtained in the theoretical calculation within the minimal
spinon mean-field model in Eq.~(\ref{spinmft}). This is because we did not
include the effect of the anisotropic spin interaction that would break
the spin rotational symmetry of Eq.~(\ref{spinmft}).
In fact, the generic spin interaction for the Yb local moments,
in the strong anisotropic limit, favors a stripe magnetic order with
the wavevectors at the M points~\cite{Yaodong}.
In a recent calculation, it was shown that the anisotropic spin interaction
enhances the spin correlation at the M points~\cite{Yaodong2}.
We note that although anisotropic spin interaction can induce a weak peak at M, the vast majority of the spectral weight is still dominated by a broad continuum across the Brillouin zone at the lowest energies measured (Fig. 2a, 3a).

\textbf{Dynamic spin structure factor of Dirac spin liquid.}

As a comparison with the spinon Fermi surface QSL,
we carry out the same calculation for the spinon mean-field Hamiltonian
with a background $\pi$-flux through each unit cell. This choice of
the background flux gives a Dirac U(1) quantum spin liquid. We fix
the gauge according to the hopping parameters that are specified
in Extended Data Fig. 6a. The spinon band structure of this mean-field Hamiltonian reads
\be
	\omega_{\bf k} = \pm {\sqrt{2}} t \sqrt{3+\cos({2k_x})+2\sin(k_x)\sin(\sqrt{3}k_y)},
\ee
where we have set the lattice constant to unity.
One observes two Dirac nodes at ${\bf k} = (\pm\frac{\pi}{2},
\mp\frac{\pi}{2\sqrt{3}})$ (Extended Data Fig. 6b), and the spinon Fermi
energy is right at the Dirac nodes.

At low energies, the only spin-1 excited states involve
either an intra-Dirac-cone spinon particle-hole pair or
an inter-Dirac-cone particle-hole pair. Therefore, the
spectral intensity of the dynamic spin structure factor
should be concentrated at the momentum transfer that
corresponds to the intra-Dirac-cone and the inter-Dirac-cone
processes. As we depict in Extended Data Fig. 6c, d, the dynamic spin
structure factor at low energies is peaked at the $\Gamma$ point,
the $M = (0,\frac{2\pi}{\sqrt{3}})$ point and the symmetry
equivalent momentum points. This result clearly differs
from the broad continuum that is observed in the experiment.
Therefore, the $\pi$-flux state, amongst other Dirac spin
liquids, are inconsistent with the experimental data.

\newpage

\renewcommand\tablename{\textbf{Extended Data Table}}
\renewcommand\figurename{\textbf{Extended Data Figure}}
\setcounter{figure}{0}

\begin{table*}[!h]
\caption{Refined structural parameters for YbMgGaO$_4$ at room temperature. Space group: $R\overline{3}m$ (No. 166). Atomic positions: Yb: 3a (0, 0, 0); Mg: 6c (0, 0, \textit{z}); Ga: 6c (0, 0, \textit{z}); O: 6c (0, 0, \textit{z})}
\begin{ruledtabular}
\renewcommand\arraystretch{0.5}
\begin{tabular}[c]{lcccc}
& a ($\AA$) & 3.40125(1) \\
& c ($\AA$) & 25.10632(16) \\
\\
\multirow{2}{*}{Yb} & \textit{B$_{11}$} ($\AA^2$) & 0.1332(18) \\
& \textit{B$_{33}$} ($\AA^2$) & 0.00204(3) \\
\\
\multirow{3}{*}{Mg} & \textit{z} & 0.21378(6) \\
& \textit{B$_{11}$} ($\AA^2$) & 0.131(4) \\
& \textit{B$_{33}$} ($\AA^2$) & 0.00161(6) \\
\\
\multirow{3}{*}{Ga} & \textit{z} & 0.21378(6) \\
& \textit{B$_{11}$} ($\AA^2$) & 0.131(4) \\
& \textit{B$_{33}$} ($\AA^2$) & 0.00161(6) \\
\\
\multirow{3}{*}{O1} & \textit{z} & 0.28887(19) \\
& \textit{B$_{11}$} ($\AA^2$) & 0.107(9) \\
& \textit{B$_{33}$} ($\AA^2$) & 0.00226(17) \\
\\
\multirow{3}{*}{O2} & \textit{z} & 0.12884(17)\\
& \textit{B$_{11}$} ($\AA^2$) & 0.137(9) \\
& \textit{B$_{33}$} ($\AA^2$) & 0.00089(18) \\
\\
& \textit{Rp} & 1.18 \\
& \textit{wRp} & 1.81 \\
& \textit{$\chi^2$} & 2.25 \\

\end{tabular}
\end{ruledtabular}
\end{table*}

\newpage

\begin{figure*}[bht]
\includegraphics{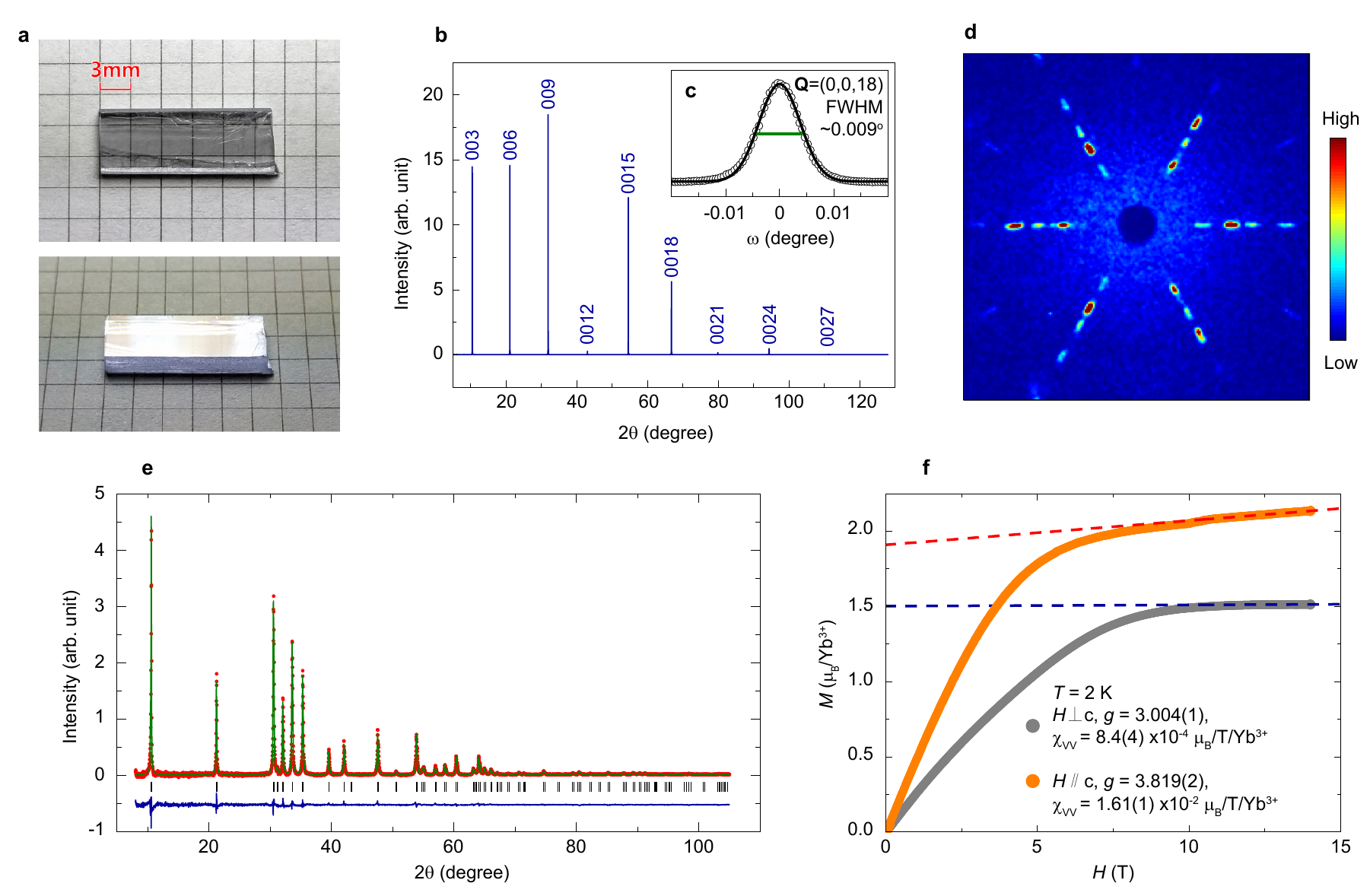}
\caption{ \textbf{Photographs, X-ray diffraction (XRD) patterns, and field dependence of magnetization of YbMgGaO$_4$.} \textbf{a,} Photographs of a representative YbMgGaO$_4$ single crystal. \textbf{b,} XRD pattern of a YbMgGaO$_4$ single crystal from the cleaved surface. \textbf{c,} Rocking curve of the (0, 0, 18) peak. The horizontal bar indicates the instrumental resolution. \textbf{d,} Laue pattern of the YbMgGaO$_4$ single crystal viewed from the \textit{c} axis. \textbf{e,} Observed (red) and calculated (green) XRD diffraction intensities of ground single crystals. The X-ray has a wavelength of $1.54$~\AA. \textbf{f,} Magnetic field dependence of magnetization at \textit{T} = 2 K. Fitted \textit{g} factors and Van Vleck susceptibility are shown in the figure.
}\label{figs1}
\end{figure*}

\newpage

\begin{figure*}[!bt]
\includegraphics{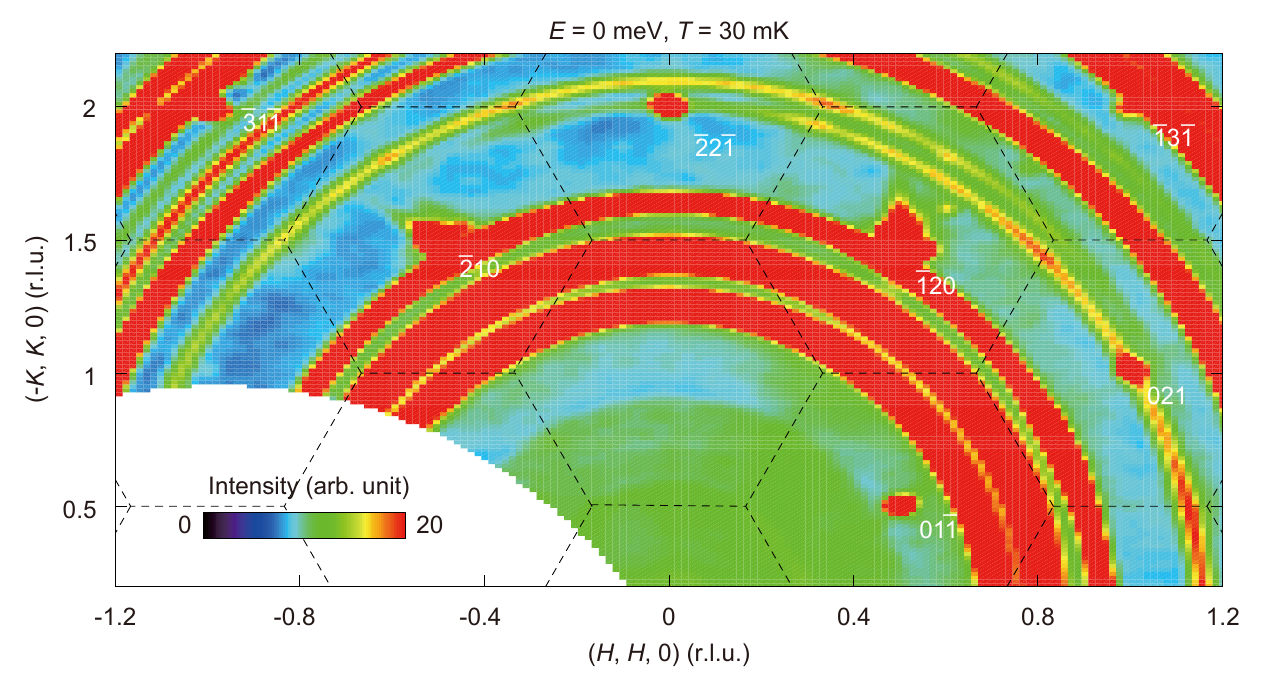}
\caption{ \textbf{Elastic neutron scattering measurements.} Elastic neutron scattering maps at ($HK0$) plane at 30 mK. No magnetic Bragg peaks were observed. The ring-like pattern is due to scattering from the polycrystalline Cu and Al sample holder. Because of the very long $c$ axis lattice constant and a small tilt of the scattering plane, some of the nuclear Bragg peak tails for $L=\pm1$ can be also seen. Dashed lines indicate the Brillouin zone boundaries.
}
\end{figure*}

\newpage

\begin{figure*}[!bt]
\includegraphics{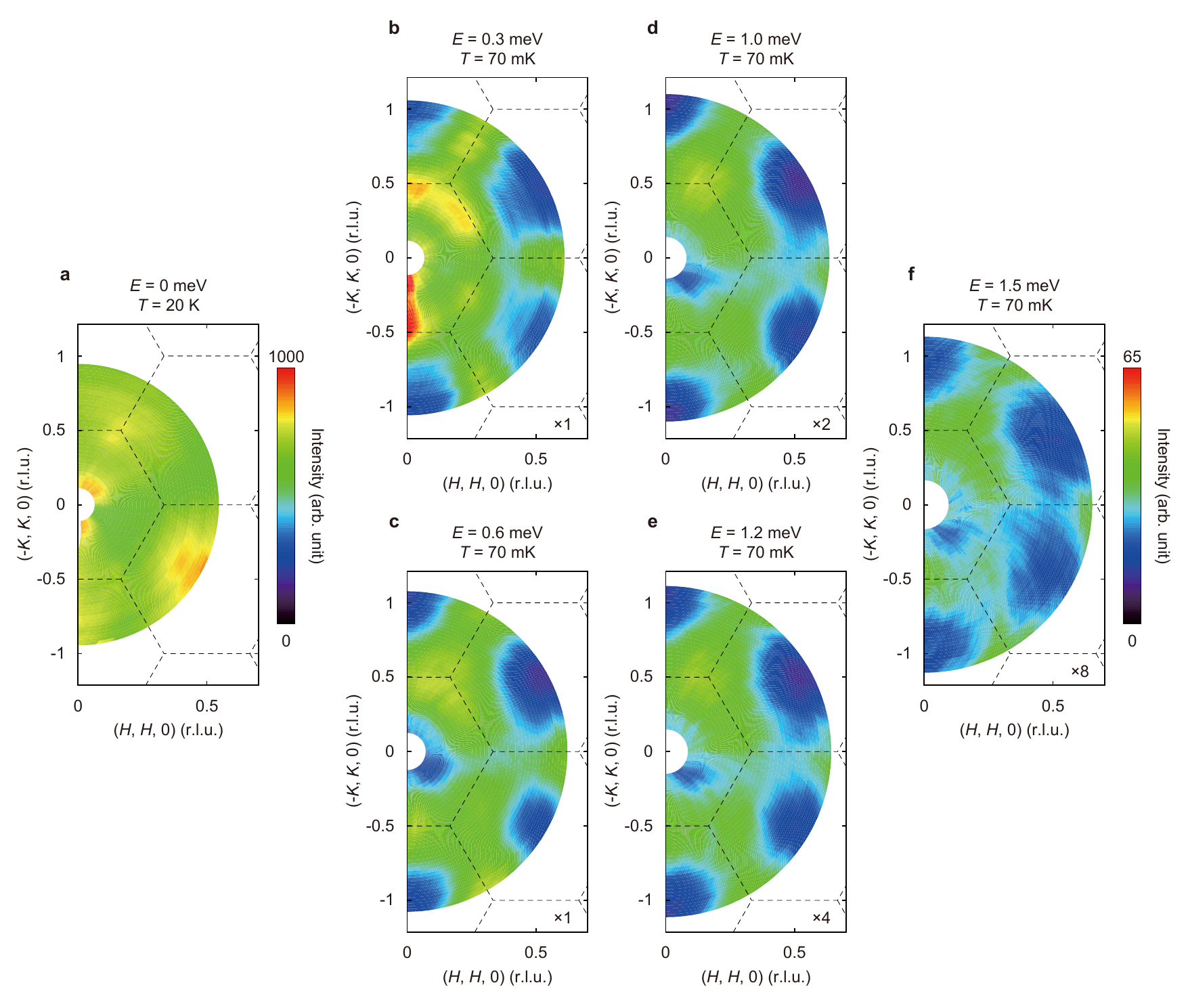}
\caption{ \textbf{Correction of neutron beam self-attenuation.} \textbf{a,} Elastic incoherent scattering image at 20 K. \textbf{b-f,} Raw constant energy images at 70 mK at indicated energies. The scattering intensity at different energies has been multiplied by a scale factor for clarity. Dashed lines indicate the Brillouin zone boundaries.
}
\end{figure*}

\newpage

\begin{figure*}[!bt]
\includegraphics{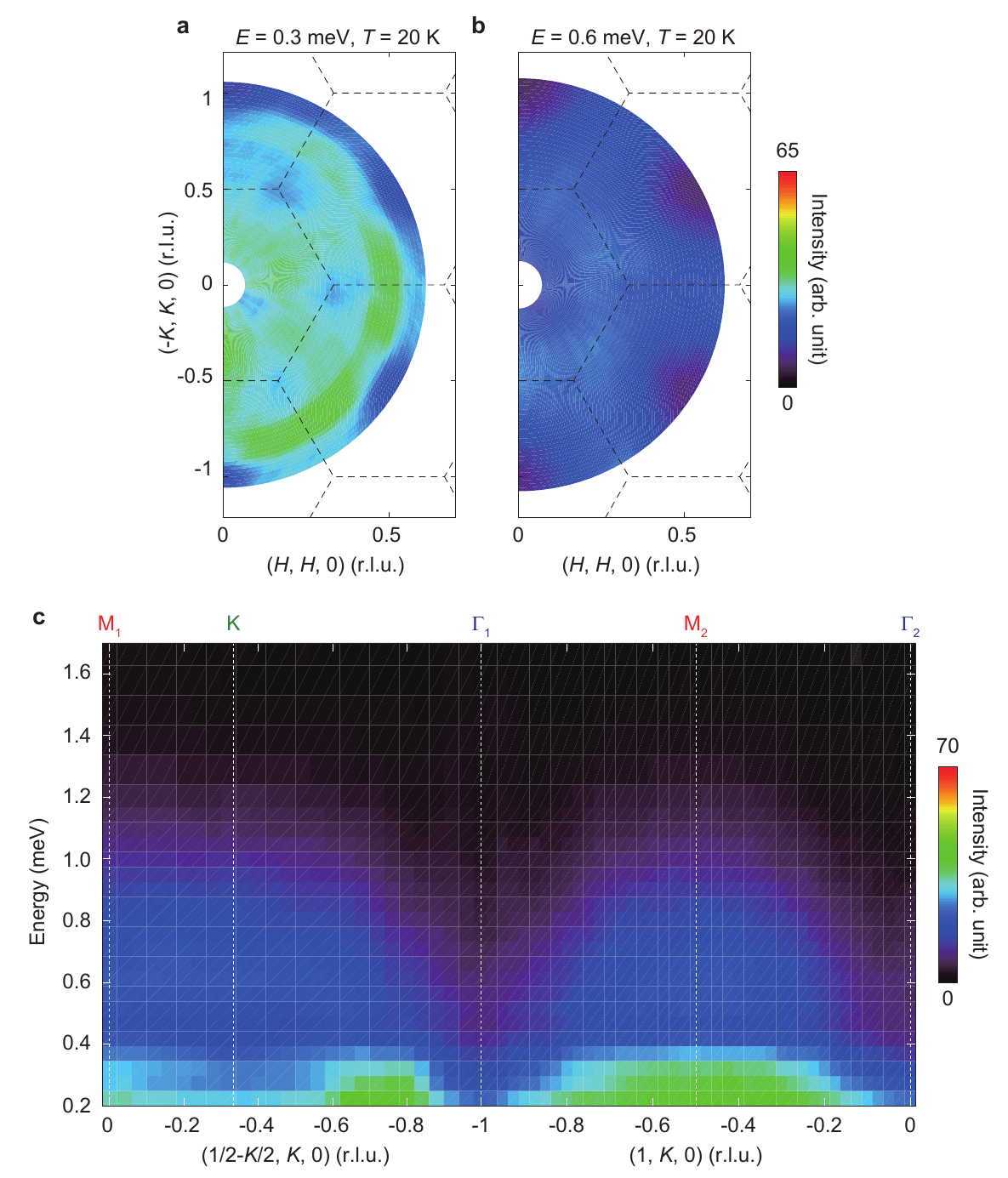}
\caption{\textbf{Additional neutron scattering data at 20 K.} \textbf{a,b,} Constant energy images at 0.3 meV and 0.6 meV at 20 K. \textbf{c,} Intensity contour plot of spin excitation spectrum along the high-symmetry momentum directions at 20 K. The scattering is broadened and weakened compared with that at 70 mK.
}
\end{figure*}

\newpage

\begin{figure*}[!bt]
\includegraphics[width=0.5\textwidth]{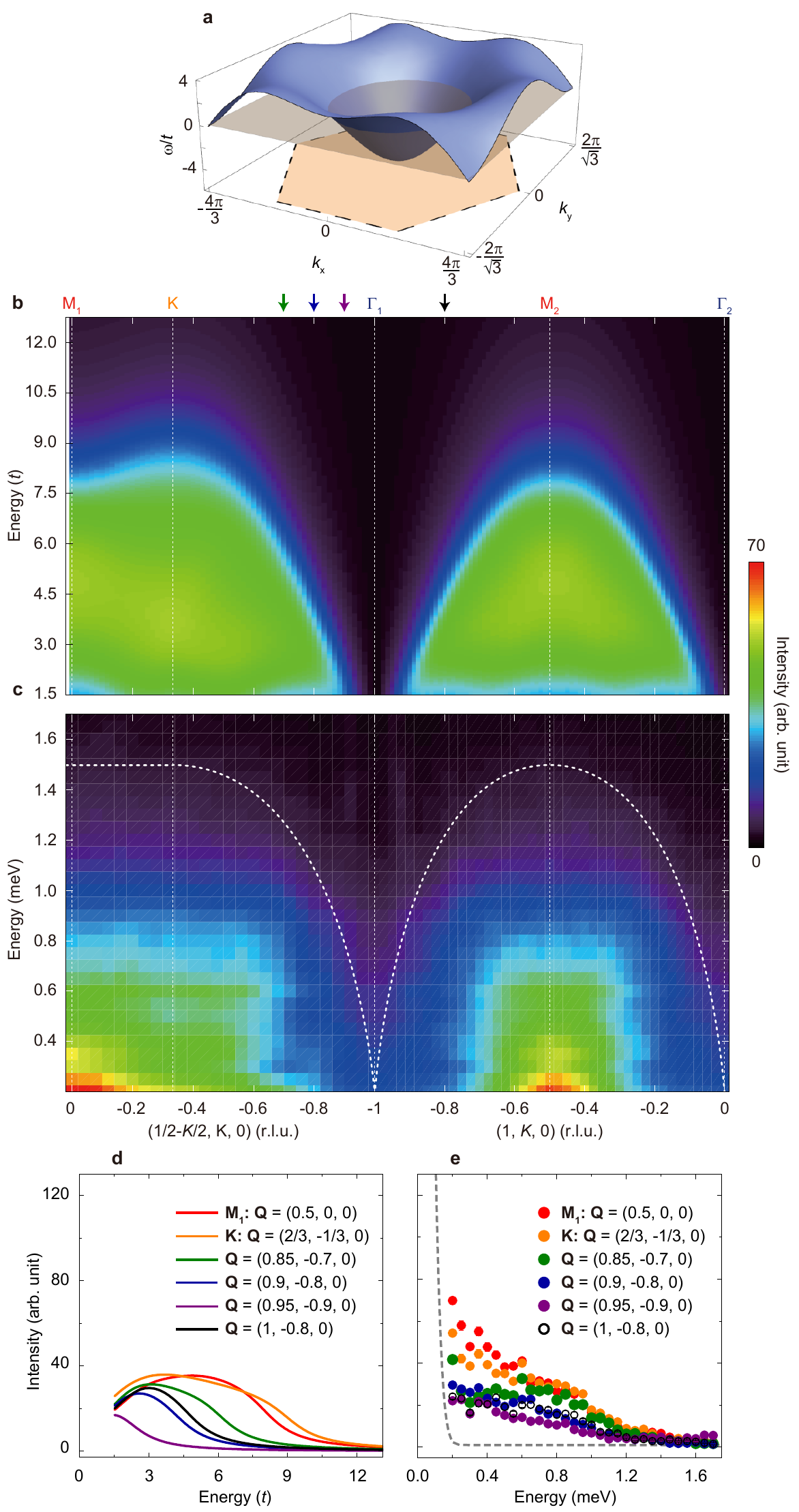}
\caption{\textbf{Calculation of 0-flux Hamiltonian.} \textbf{a,} Spinon dispersion $\omega_{\bf k}$ of the 0-flux Hamiltonian. The grey plane marks the Fermi level at $\omega = 0$, and its intersection with the band gives the Fermi surface. The light orange hexagon represents the projection of first Brillouin zone. The maximum of $\omega_{\bf k}$ is $3t$ whereas the minimum is $-6t$, providing a bandwidth of $9t$. \textbf{b,} Calculated dynamic spin structure factor along high symmetry directions. A reciprocal lattice unit (r.l.u.) is used here, which can be obtained by $H = \frac{k_x}{4\pi}-\frac{\sqrt3k_y}{4\pi}, K = \frac{k_x}{4\pi}+\frac{\sqrt3k_y}{4\pi}$. \textbf{c,} Measured spin excitation spectrum along high symmetry directions at 70 mK. \textbf{d,} Calculated energy dispersion at indicated momenta (marked by arrows in \textbf{b}). \textbf{e,} Measured constant $\bf{Q}$ scans at indicated momenta. The dashed line is the incoherent elastic line for $E_f=4$ meV.
}
\end{figure*}

\begin{figure*}[!bt]
\includegraphics{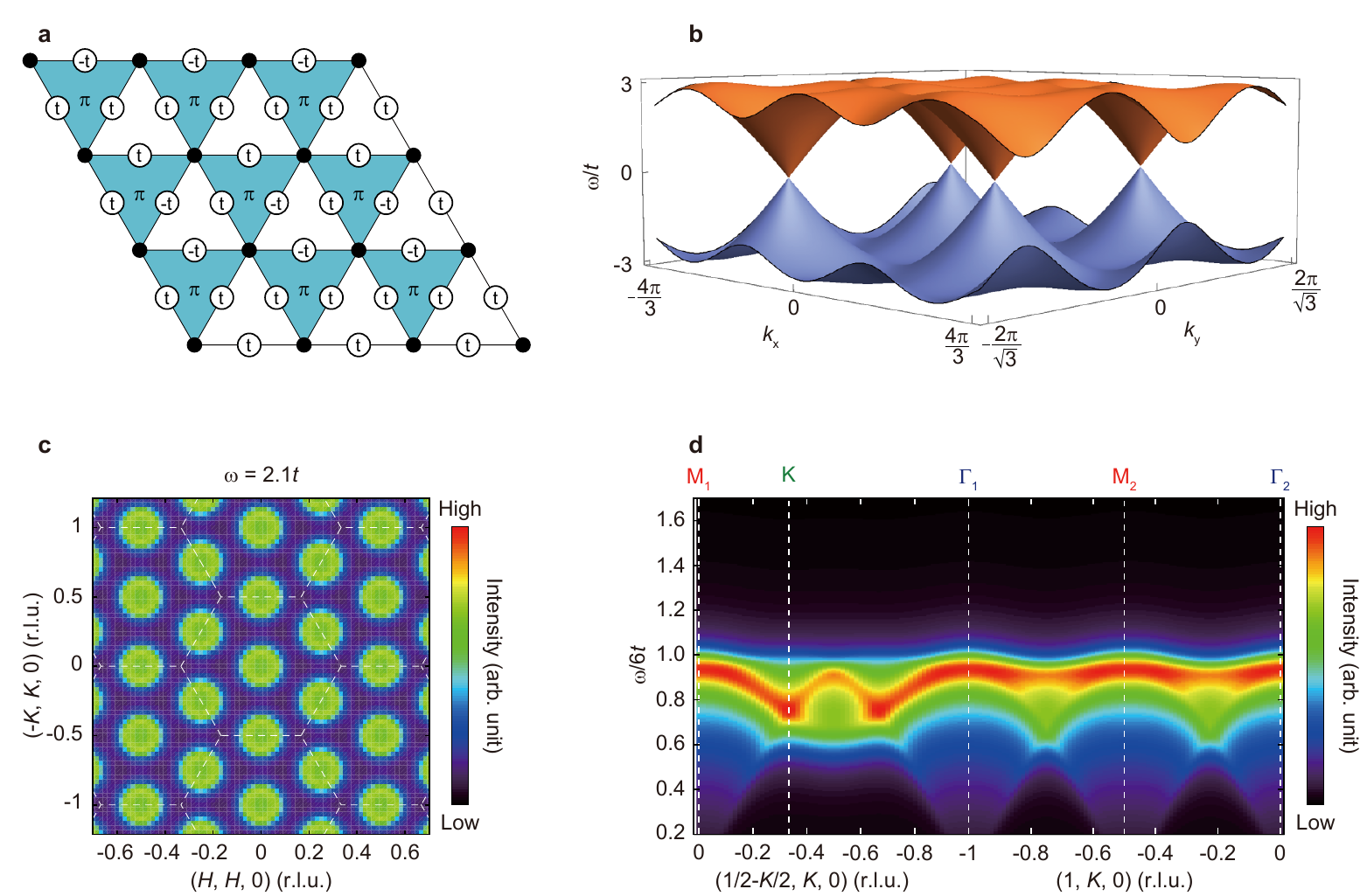}
\caption{\textbf{Calculation of $\pi$-flux Hamiltonian.} \textbf{a,} Flux pattern and real nearest neighbor hoppings on the triangular lattice. In the figure, ``$+t$'' denotes $t_{ij} = t_{ji} = t$ and ``$-t$'' denotes $t_{ij} = t_{ji} = -t$. Moreover, $\pi$'s denote triangles which are threaded by a $\pi$ flux. \textbf{b,} Spinon band structure of the $\pi$-flux Hamiltonian. The two bands are particle-hole related, both with bandwidths of $3t$. \textbf{c,} Calculated momentum dependence of the dynamic spin structure factor at low energy $\omega = 2.1t$. Strong peaks can be distinguished at the $\Gamma$ point, the M = $(0,\frac{2\pi}{\sqrt{3}})$ point [$(\frac{1}{2}, -\frac{1}{2})$ in r.l.u.] and equivalent positions. White dashed lines denote for the zone boundaries. \textbf{d,} Calculated dynamic spin structure factor along high symmetry points with $\eta = 0.3t$.
}
\end{figure*}

\end{document}